\documentclass{aa}
\usepackage{graphics}
\usepackage{color}
\usepackage{natbib}
\newcommand{\cmtwo}{cm$^{-2}$}  
\newcommand{\cmthree}{cm$^{-3}$}

\newcommand{\kms}{km\,s$^{-1}$}       %km/s
\newcommand{\um}{$\mu$m}                                 %micron
\newcommand{\molh}{H$_{2}$}                              %H_2, H_2O and H II

\newcommand{\water}{H$_{2}$O}

\newcommand{\lapprox}{$\stackrel {<}{_{\sim}}$}
\newcommand{\about}{$\sim$}                       %approx
\newcommand{\ctwo}{[C\,{\sc ii}]\,157\,$\mu$m}    %fine structure lines 

\newcommand{\oishort}{[O\,{\sc i}]\,63\,$\mu$m}
\newcommand{\oilong}{[O\,{\sc i}]\,145\,$\mu$m}

\newcommand{\av}{$A_{\rm V}$}                     %extinction

\newcommand{\asec}{$^{\prime \prime}$}

\begin{document}

   \title{Outflows from young objects observed with the ISO-LWS\thanks{Based
          on observations with ISO, an ESA project with instruments
          funded by ESA Member States (especially the PI~countries:
	  France, Germany, the Netherlands and the United Kingdom) and with the
	  participation of ISAS and NASA.}
  }

   \subtitle{I. Fine structure lines \oishort, \oilong\ and \ctwo}

   \author{     R.\,Liseau\inst{1}	  \and
		K.\,Justtanont\inst{1}   \and  	 
		A.G.G.M.\,Tielens\inst{2} 
	}

   \offprints{R. Liseau}

   \institute{ Stockholm Observatory, AlbaNova University Center, Roslagstullsbacken 21, SE-106 91 Stockholm, Sweden  
   \and 
  	      Kapteyn Astronomical Institute, P.O. Box 800, 9700 AV Groningen, Netherlands \\
   	\email{rene.liseau@astro.su.se, kay@astro.su.se and tielens@astro.rug.nl }
   }

\date{Received date: \hspace{5cm}Accepted date:}

   \abstract{Far infrared fine structure line data from the ISO archive have been extracted for several hundred YSOs and their outflows, including molecular (CO) outflows, optical jets and Herbig-Haro (HH) objects. Given the importance of these lines to astrophysics, their excitation and transfer ought to be investigated in detail and, at this stage, the reliability of the diagnostic power of the fine structure transitions of O\,I and C\,II has been examined. Several issues, such as the extremely small intensity ratios of the oxygen 63\,\um\ to 145\,\um\ lines, are still awaiting an explanation. It is demonstrated that, in interstellar cloud conditions, the 145\,\um\ line is prone to masing, but that this effect is likely an insufficient cause of the line ratio anomaly observed from cold dark clouds. Very optically thick emission could in principle also account for this, but would need similar, prohibitively high column densities and must therefore be abondoned as a viable explanation. One is left with \oishort\ self absorption by cold and tenuous foreground gas, as has been advocated for distant luminous sources. Recent observations with the submillimeter observatory Odin support this scenario also in the case of nearby dark molecular clouds. On the basis of this large statistical material we are led to conclude that in star forming regions, the [O\,I] and [C\,II] lines generally have only limited diagnostic value. 

               }
   \maketitle

%
%________________________________________________________________

\section{Introduction}

In the optically thin limit, the far infrared lines of [O\,I] and [C\,II] provide valuable diagnostics for the densities and temperatures of the emitting regions. This is the case, when the levels are excited by collisions with electrons and/or atomic hydrogen in hot to warm interstellar regions, generating major thermal energy sinks. This cooling radiation is so powerful that it is readily detectable over intergalactic distances. Since these transitions are at long wavelengths, they do not suffer from extinction in the dust enshrouded regions of ongoing star formation.

The heating of these regions could reflect the action of powerful shocks driven by jets and outflows from young stellar objects \citep{hollenmckee1989}. Alternatively, FUV photons from nearby young massive stars can photodissociate and heat surrounding gas, forming a PhotoDissociation Region \citep[PDR;~][]{th1985}. Both of these types of sources are known to be bright sources of infrared atomic fine structure line emission. It is thus instrumental to understand the details of the energy (and momentum) injection into the interstellar gas, if one wants to understand the emission and transfer of this radiation. In general, this involves a set of complex assumptions and the need to specify a large number of parameters, the precise values of which are often not well known. As a result, basic processes may be concealed or difficult to recover. 

For these reasons, we will analyze a clean and simple case, where we assume that the energy is locally available as heat, without paying initially much attention to how the energy has been generated, delivered and dissipated. In this first contribution, which is observational and statistical in nature, we focus on the emission in the fine structure lines of O\,I and C\,II from outflows in regions of star formation. The statistics should disclose significant trends regarding the emission processes, comprehensible with a minimum of assumptions in simple models of the atomic physics of these simple transitions. A more sophisticated analysis, possible only for a limited number of individual sources with known distances, will be deferred to a future paper. 

The paper is organized as follows: In Sect.\,2, we describe the data selection and in Sect.\,3, we present the results. These are discussed in Sect.\,4, where we first address their reliability in contrast to mere spurious instrumental effects. We then examine various excitation models aimed at explaining the observations, initially using purely analytical methods, giving exact solutions, but further on also numerical approaches. Finally, in Sect.\,5 we briefly summarize our main conclusions. 

\section{Source selection and data reduction}

\subsection{ISO archive}

Although the Infrared Space Observatory (ISO) was decomissioned several years ago in 1998, much of the data collected during the mission still need proper evaluation and analysis. Thanks to the availability of and relatively easy access to the ISO data 
archive\footnote{\texttt{http://www.iso.vilspa.esa.es/ida/index.html}} these demands can be met in practice. Of the retrieved LWS data files \citep[Long~ Wavelength~ Spectrometer;~][]{clegg1996}, each contains spectral scans from roughly 45\,\um\ to 200\,\um\ for a particular position in the sky. The search key words (Scientific Category) were {\it Interstellar Matter, Stellar Physics: bipolar outflows, circumstellar disks, star formation, young stellar objects}, which resulted in a total of 552 LWS medium resolution observations in the data base. Spatial {\it maps} are counted as {\it one observation}, containing the observations of several positions in the sky (thus totalling 743).

The reduction of these data is based on the Off Line Processing version 10 (OLP\,10), resulting in fully calibrated spectral data sets. For \oishort\ and $145$\,\um\ and \ctwo, the spectral line fluxes were extracted by Gaussian fitting of the spectral flux density at the line position. The error estimates are based on half the peak-to-peak noise in the surrounding continuum and are, as such, conservative. Saturated data were, of course, disregarded altogether. Limits are given as $3\sigma$ and have been obtained as $3 \times {\rm rms} \times {\rm spl}$, where rms corresponds to the continuum flux density at the wavelength of the line and where the spectral resolution element spl = 0.29\,\um\ for the 63\,\um\ line and spl = 0.6\,\um\ for the 145\,\um\ and 157\,\um\ lines, respectively.

\section{Results}

\subsection{Comparison with other work}

\begin{figure}[t]
  \resizebox{\hsize}{!}{
  \rotatebox{00}{\includegraphics{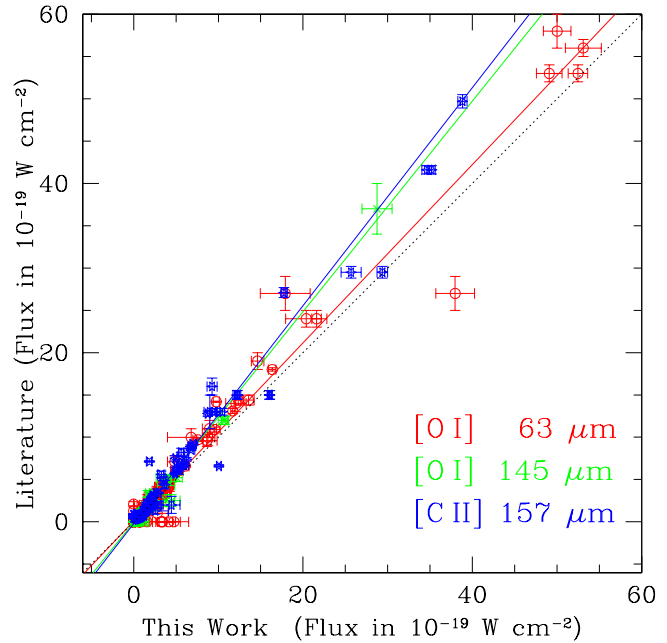}}
  }
  \caption{Comparison of line flux values obtained in this work to those found in the literature. For clarity, only the initial section of the distributions is shown (0 to $60 \times 10^{-19}$\,W\,\cmtwo), but the linear regression fits refer to the entire data sets (up to $400 \times 10^{-19}$\,W\,\cmtwo). The dotted line is of unit slope and meant to guide the eye only. Next
to this line is the fit for \oishort, followed by that for \oilong\ and \ctwo, respectively.
  }
  \label{kay_lit}
\end{figure}

A subset containing almost 100 ISO pointings was used to check for systematic trends in our measuring procedure. For these objects, line fluxes have been published and were obtained by a number of different individuals, ensuring randomisation. Fig.\,\ref{kay_lit} shows this comparison for the three fine structure lines \oishort, \oilong\ and \ctwo, from which it is apparent that systematic trends are indeed present. Statistically, our line fluxes tend to be lower than the literature values, and this trend appears more pronounced for the \oilong\ and \ctwo\ lines than for the \oishort\ line, with deviations from unity by about 25\% and 5\%, respectively (Fig.\,\ref{kay_lit}). 

The reason for this can be understood as a `maturity' effect, in the sense that the behaviour of the detectors became known better with time. The literature data refer all to OLP versions 7 or earlier, i.e. to versions in which the final LWS calibration had not yet been established. The archive data, on the other hand, are uniformly based on OLP\,10, where the Relative Spectral Response Function (RSRF) is higher compared to earlier versions. Since the spectral flux density is inversely proportional to the RSRF, a lower value results for a given photo current. This effect is particularly pronounced in the LWS bands of the 145\,\um\ and 157\,\um\ lines. At the end of the mission, the LWS seemed thus more sensitive. However, the refinement of these RSRFs remained within the absolute calibration accuracy of 30\% announced already early in the mission \citep{swinyard1996}.

\subsection{Observed line ratios}

\begin{figure}[t]
  \resizebox{\hsize}{!}{
  \rotatebox{00}{\includegraphics{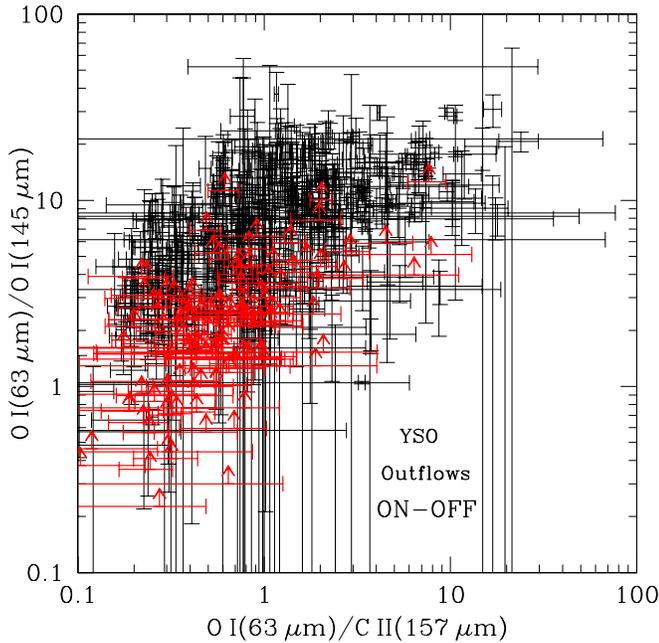}}
  }
  \caption{Line ratio plot for the fine structure lines of O\,I and C\,II. Dots with error bars designate ratios based on differential line fluxes (ON-OFF measurements), whereas the arrows indicate lower limits at the $3 \sigma$ level. In the majority of observations, an OFF-source measurement was not made and in several cases, the OFFs refer to the average flux in the associated flows or, for non-mapped objects, the symbols refer merely to ON source measurements (see the text).
  }
  \label{obsratios_onoff}
\end{figure}

The ratios of the intensities of the fine structure transitions of neutral oxygen and ionized carbon are plotted for outflows and their sources in Fig.\,\ref{obsratios_onoff}. The upper bounds of a ${\rm few} \times 10$ of both ratios most likely reflect the limited dynamical range of the LWS.

Many symbols with assigned error bars in Fig.\,\ref{obsratios_onoff} identify flux differences (ON-OFF) prior to the ratioing, where the OFF refers to a reference position sufficiently distant from the source to contain only background radiation, if any. However, many of the data lack such a proper OFF-source measurement. In those cases where the sources had been mapped, it was generally found that the intensities in the map were much lower than that of the ON position. Consequently, an average flux of the
map, excluding the ON position, is used instead. Finally, the majority of sources has only an ON measurement. In summary, the points in Fig.\,\ref{obsratios_onoff} represent a heterogeneous data set, with the potential risk that the quality could be insufficient for a statistical analysis.

\begin{figure}[t]
  \resizebox{\hsize}{!}{
  \rotatebox{00}{\includegraphics{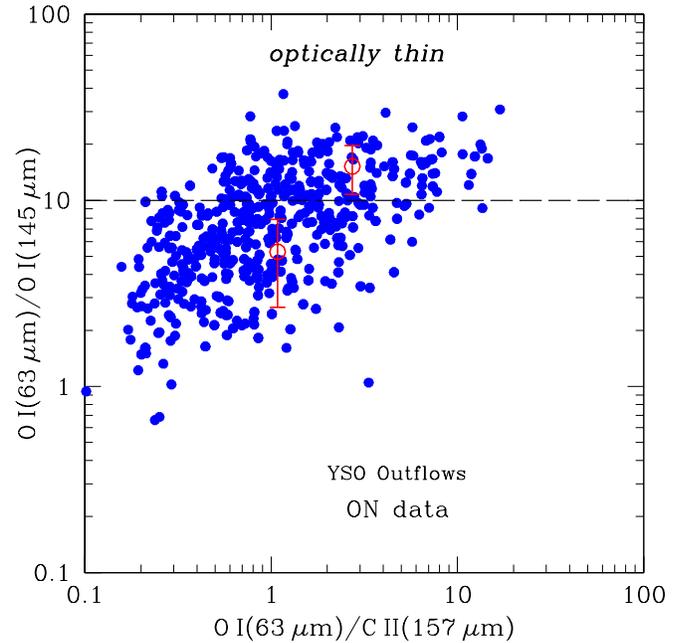}}
  }
  \caption{Similar to Fig.\,\ref{obsratios_onoff} but only ON-source measurements are shown, i.e. no subtraction of OFF or off data have been made. Limit symbols and error bars have been omitted to avoid crowding (see Fig.\,\ref{obsratios_onoff}). The horizontal dashed line indicates the optically thin regime for the [O\,I]\,63\,\um/[O\,I]\,145\,\um\ line ratio above about ten. The two open symbols with vertical error bars refer to the average values of data lying above and below the dividing line, respectively.
  }
  \label{obsratios}
\end{figure}

Selecting only one category of data will always introduce the same type of error, deminishing biasing effects on the sample. For example,
the [O\,I] background is observed to be generally very weak (because of the relatively high excitation). Rather than leading to good data quality, the subtraction of an OFF (or off) flux could introduce unwanted extra noise into the ON-OFF/ON-off pair. It might, therefore, be justified to use {\it only} ON data for the diagram and judge the quality a posteriori. Because \ctwo\ emission is found almost everywhere in the sky, the corresponding \oishort/\ctwo\ will be an under-estimate and for this and other reasons (Sect.\,4.6), results involving \ctwo\ will only marginally be considered in this paper. This analysis would have to rely on self-consistent model calculations  and will thus be deferred to a future paper.

In Fig.\,\ref{obsratios}, line ratios based on only ON data are shown. Obviously, the two distributions shown in Fig.\,\ref{obsratios_onoff} and \ref{obsratios}, respectively, are {\it statistically} indistinguishable. For values of the ratio \oishort/\oilong\,$>10$, the emission is likely optically thin \citep[see~below~and,~e.g.,][]{th1985}. Taking \oishort/\oilong\,=\,10 as a dividing line yields the averages $(2.7 \pm 3.0,\,15.2 \pm 4.4)$ above and $(1.1 \pm 1.3,\,5.3 \pm 2.6)$ below, respectively. These averages are shown as open symbols with vertical bars in Fig.\,\ref{obsratios} and they will find their application in Sect.\,4.5. 
 
As expected (and demonstrated), the ratio \oishort/\oilong\ remains essentially unaffected, when one neglects the OFF source subtractions. Therefore, the major observational result of this paper, namely that merely 35\% of the data have ratios $\ge 10$, rests on a solid foundation and forms the basis for the following discussion. 

\section{Discussion}

\subsection{Instrumental effects}

\subsubsection{Detector responsivities}

Of interest here is of course the intercalibration among different detectors. All three lines fall onto GeGa photoconductors, with the \oilong\ and \ctwo\ lines falling onto stressed detectors.  The relative trends of the RSRFs likely reflect also the behaviour of these detectors in an absolute sense and any instrumental origin for small oxygen line ratios could be safely disregarded: relative to \oishort, the \oilong\ line would get assigned a lower flux than it should have, which would lead to an {\it over-}estimate of the $I_{63}/I_{145}$ line ratio, in contrast to the observed behaviour. 

As can be seen in Fig.\,\ref{obsratios}, the \oishort/\oilong\ and \oishort/\ctwo\ data have different lower cut-offs, with the latter filling the region between 0.1 and 1, whereas the former do not. The \oilong\ and \ctwo\ lines both fall onto the same detector with similar RSRFs. The cause of the differences of Fig.\,\ref{obsratios} is thus not instrumental in nature. We will return to this point in Sect.\,4.6.

\subsubsection{LWS beam sizes}

Estimated sizes of the non-diffraction-limited LWS beam appear non-uniform with wavelength \citep{gry2003}. It is conceivable, therefore, that the observed line ratios in Figs.\,\ref{obsratios_onoff} and \ref{obsratios} could be spuriously affected by this. However, in both cases, the effect would be less than a factor of 1.5, which would not be sufficient to explain any line ratio anomaly in excess of one order of magnitude.

\subsection{Theoretical line ratios}

\begin{figure}[t]
  \resizebox{\hsize}{!}{
  \rotatebox{00}{\includegraphics{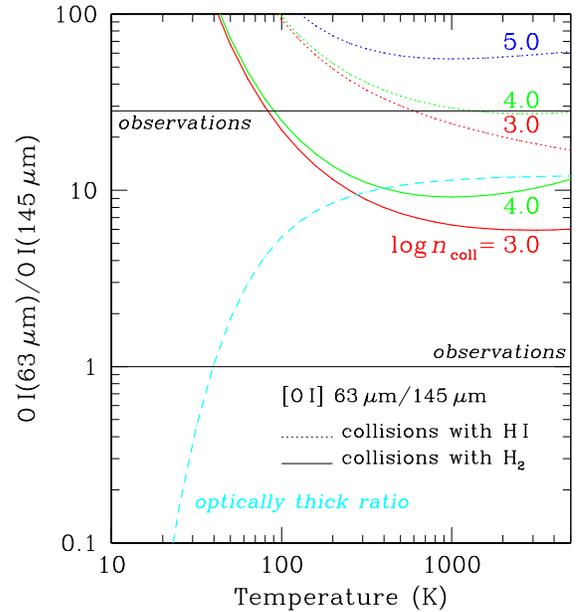}}
  }
  \caption{\oishort/\oilong\ line intensity ratios for collisions with neutral hydrogen, H\,I, as the function of the temperature are shown by the dotted lines. The logarithms of the densities (in \cmthree) are indicated above each curve. Optically thin ratios are given for the volume densities $n({\rm H})= 10^3$, $10^4$ and $10^5$\,\cmthree. The broken line outlines the ratio of optically thick lines of equal velocity width, each of which is approximated by the integrated Planck function at the given temperature, i.e.$\int_{\Delta \nu}\!I_{\nu}\,d\nu = \Delta \nu B_{\nu}(T)$. The optically thin intensity ratios for the inverted levels, due to collisions with \molh, are shown by the solid lines, with the logarithmic densities in \cmthree\ below. Observed line ratios fall roughly into the area inside the thin horizontal lines (compare with Fig.\,\ref{obsratios}). 
  }
  \label{theoratios}
\end{figure}

Theoretical estimates of line intensity ratios can be obtained from excitation calculations, i.e. from the solutions to the statistical equilibrium equations (see the Appendix). Comparing these to the observed \oishort/\oilong\ line ratios indicates that optically thin emission seems essentially ruled out for most temperatures and densities (see the dotted curves in Fig.\,\ref{theoratios}). 

The observations could be reconciled with optically thick emission, as indicated by the dashed curve in the figure. Temperatures would have to be low though, below 100\,K. There are at least two difficulties with this scenario.

\begin{figure}[t]
  \resizebox{\hsize}{!}{
  \rotatebox{00}{\includegraphics{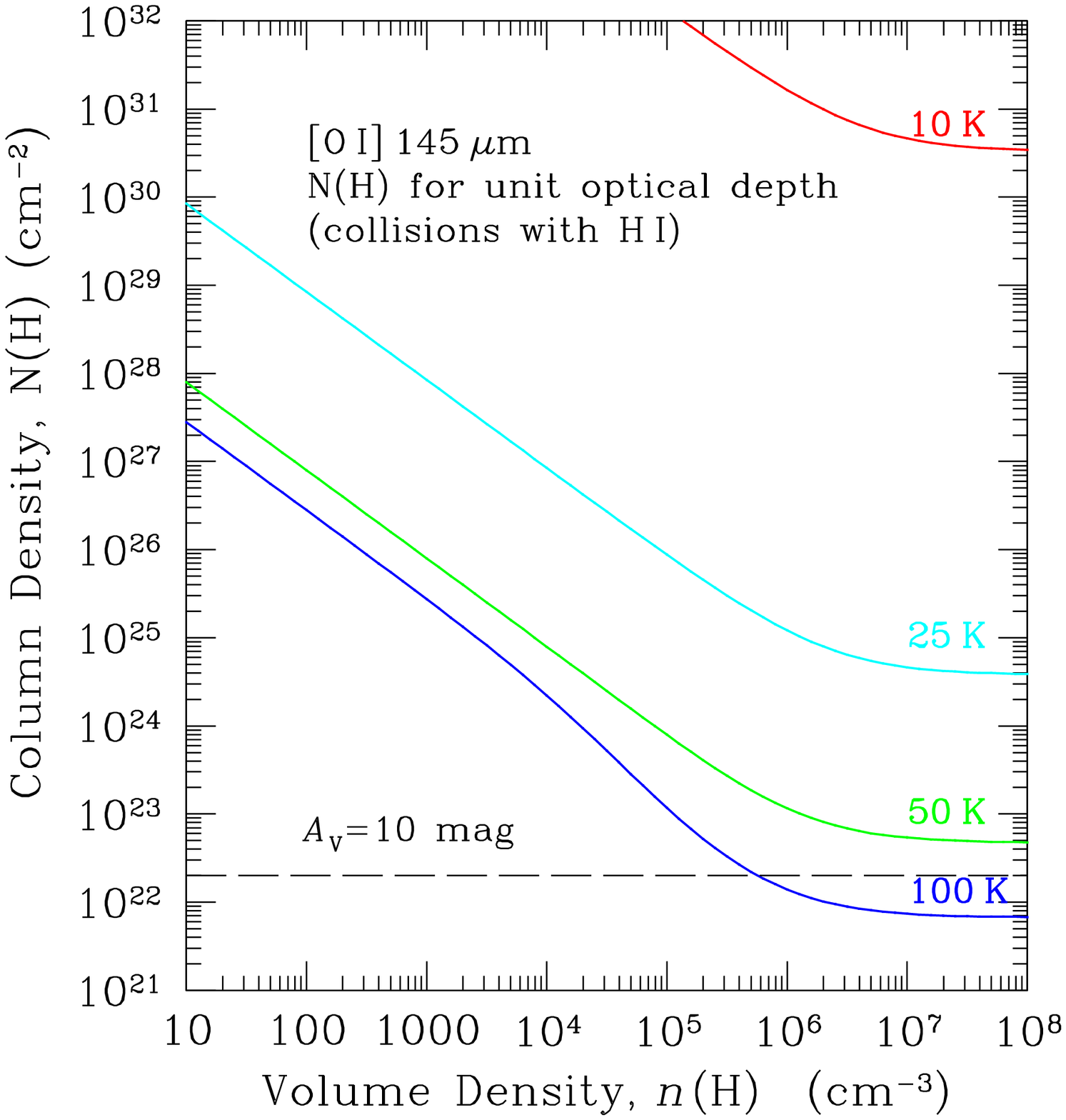}}
  }
  \caption{Hydrogen column densities $N({\rm H})$ (in \cmtwo) versus volume densities $n({\rm H})$ (in \cmthree) for unit line center optical depth in the \oilong\ line, $\tau_0=1$. Collisional excitation with atomic hydrogen is assumed, in addition to a cosmic oxygen abundance and a Gaussian line width of 1\,\kms. At temperatures below 100\,K, column densities for optically thick \oilong\ emission become prohibitively large.
  }
  \label{coldens}
\end{figure}

Firstly, the locus for the optically thick ratios has been computed for {\it both} lines being optically very thick (the line source functions approach the Planck functions). Not unexpectedly, at the low temperatures characteristic of dark clouds, the 145\,\um\ line would require excessive column densities, hardly ever encountered in molecular clouds (Fig.\,\ref{coldens}).

Secondly, detailed and self-consistent models of either shocks or PDRs, having `acceptable' total column densities, do not predict such low values of the FIR [O\,I] line ratios over the entire feasible parameter space \citep[e.g.,~][]{kaufman1999}. An extensive discussion can be found in \citet{liseau1999}. 

\subsection{Inversion of the level populations}

The line ratios in Fig.\,\ref{theoratios} have been computed for oxygen atoms colliding with hydrogen atoms \citep{Launey1977}. At the low temperatures of the optically thick locus, hydrogen is expected to be mostly in molecular form, however. In this case, collisional excitation of the [O\,I] levels will be with \molh\ (and He). The $^3P_J$ ground state of O\,I has energies $E_J$, such that $E_2 < E_1 < E_0$, each with statistical weight $g_J = 2\,J + 1$. The 63\,\um\ line emission originates from transitions $E_1 \rightarrow E_2$, while
the 145\,\um\ line is due to $E_0 \rightarrow E_1$.

Collisional transitions $J=0 - 1$ are forbidden to first order \citep{monteiro1987}, which together with the fact that the Einstein transition probability $A_{01} < A_{12}$ leads to the possibility of inverted level populations, $n_J$ \citep[see also][]{smith1969}. In this case, the inversion parameter is

\begin{equation}
\frac {g_0}{g_1} - \frac{n_0}{n_1} < 0 
\end{equation}

i.e. would become negative at low densities ($n_{\rm coll} \ll n_{\rm crit}$) and turn over to positive values toward the high density thermalization regime ($n_{\rm coll} \gg n_{\rm crit}$), where the critical density is defined as $n_{\rm crit}=A_{01}/q_{01}(T)$ with the denominator being the collision rate coefficient.

\begin{figure}[t]
  \resizebox{\hsize}{!}{
  \rotatebox{00}{\includegraphics{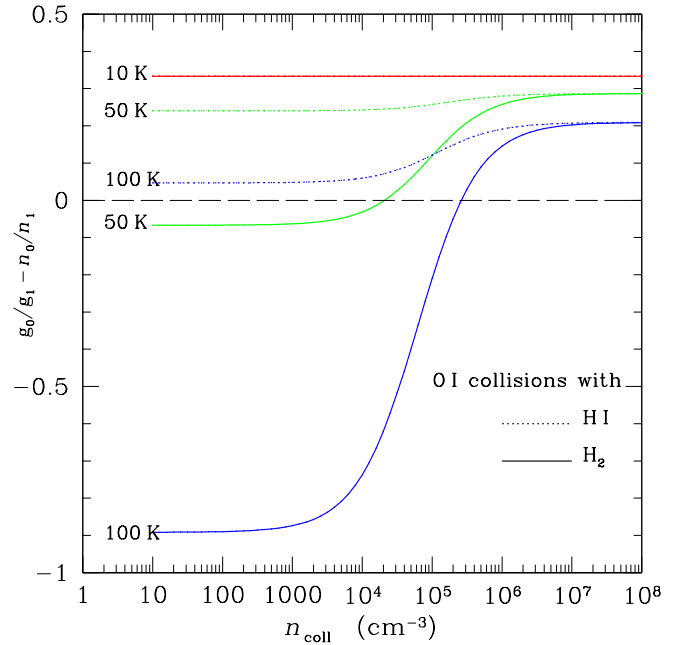}}
  }
  \caption{The level inversion parameter $g_0/g_1-n_0/n_1$ for O\,I is shown for collisional excitation with atomic hydrogen (dotted lines) and molecular hydrogen (solid lines), respectively. For collisions with \molh, this parameter is negative at temperatures above 50\,K and densities below $10^5$\,\cmthree, implying inversion of the upper level population. At low temperatures, this parameter becomes independent of the density and the two curves become indistiguishable, approaching the ratio of the statistical weights ($g_0/g_1-n_0/n_1\rightarrow g_0/g_1 = 1/3$, see the curves for 10\,K).  
  }
  \label{inversion}
\end{figure}

In Fig.\,\ref{inversion}, the inversion parameter for the oxygen levels $J=0$ and 1 is shown as a function of the density for collisions with H\,I and \molh, respectively, and for three values of the temperature. Whereas $g_0/g_1-n_0/n_1 > 0$ always for collisions with H\,I, this parameter is seen to be negative for \molh\ collisional excitation at temperatures exceeding a few times ten degrees and at densities lower than the critical density. Optically thin line intensity ratios are then expected to become different too. 

The collisional rates were adopted from \citet{Jaquet1992} and computed for a thermally weighted distribution of ortho- and para-\molh, implying preferential collisions with para-\molh\ at low temperatures. For $T \le 10^3$\,K, collisions with ortho-\molh\ are about two times faster. Hence, in an ortho-dominated gas, the magnitude of the level inversion would become somewhat reduced.  

Helium makes up about 20\% of the molecular gas by number, but including collisions with He \citep{monteiro1987} alters the O\,I results only marginally. Similar is true for the He contribution to the excitation in an H\,I dominated gas.

\subsubsection{Comparison with the observations}

Fig.\,\ref{theoratios} shows \oishort/\oilong\ as a function of the temperature for collisional excitation with either H\,I or \molh\ (the dotted and solid curves, respectively). As expected, collisions with \molh\ yield smaller line ratios, in closer agreement with the observations. At densities $n({\rm H}_2) < 10^5$\,\cmthree, the loci of these optically thin ratios are crossing that of the optically thick case. 

This dependence on the exact character of the collisional excitation process reduces the diagnostic value of the oxygen lines. For example, a line ratio \oishort/ \oilong\,\about\,20 could indicate either molecular gas at 100\,K or atomic gas at 2\,000\,K (although the line intensities would be very different). Another degeneracy exists for a ratio of about ten, with either optically thin or thick emission as apparently viable, but contradictory, options. 

\subsubsection{Masing atomic transitions}

Level inversion implies potentially laser or maser activity. The opacity at the line center $\lambda_0$ is given by

\begin{equation}
\kappa(v_0)= \frac {\ln^{1/2}2}{4\,\pi^{3/2}} \frac{\lambda_0^3}{\Delta v_{\rm fwhm}}
 A_{01}\,a({\rm O})\,n({\rm H})\,f_1 \left ( \frac {g_0}{g_1} - \frac {n_0}{n_1} \right )
\end{equation}

where a Gaussian profile with full width at half maximum $\Delta v_{\rm fwhm}$ and center velocity $v_0$ has been assumed; $a({\rm O})$ is the abundance of oxygen relative to hydrogen and the other symbols have their usual meaning. With the inversion parameter being negative, the optical depth in the line becomes negative, $\tau_0 = \kappa(v_0)\,Z <0$, where $Z$ is the physical length of the column along the line of sight. For amplified background radiation, $I = I_0\,e^{-\tau_0}$, a masing 145\,\um\ line could lead to even lower line ratios than indicated in Fig.\,\ref{obsratios} and potentially explain the very low observed values. 

\begin{figure}[t]
  \resizebox{\hsize}{!}{
  \rotatebox{00}{\includegraphics{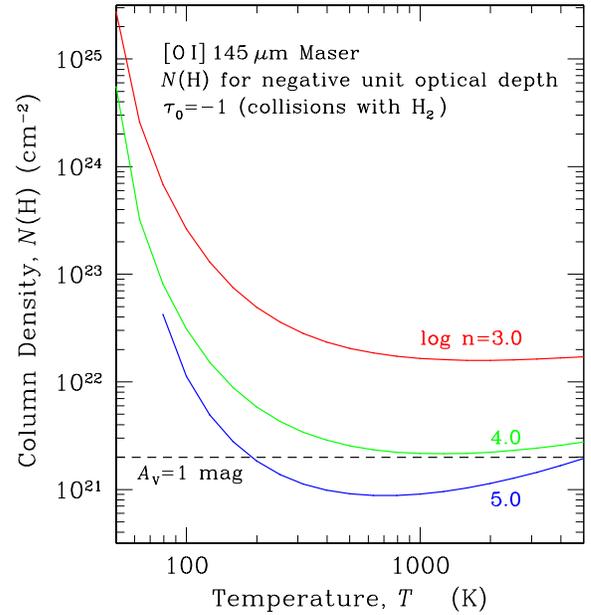}}
  }
  \caption{The hydrogen column density, $N({\rm H})$, of a parcel of molecular cloud gas along the line of sight to attain an optical depth of $\tau_0=-1$ in a thermally broadened 145\,\um\ maser line. Logarithmic densities in \cmthree\ are given next to each curve.  
  }
  \label{gain}
\end{figure}

The smallest possible column density $N({\rm H})=n({\rm H})\,Z$ for a gain by a factor of $e^1$, i.e. for $\tau_0=-1$ at line center, is obtained for a minimal, i.e. thermal, line width and this case is shown in Fig.\,\ref{gain} for a few typical cloud densities. At low temperatures, characteristic of dark clouds, these columns are clearly in excess of what could be found in such objects. There, masing seems thus not capable of explaining any hypothetical 145\,\um\ excess. 

This situation may change in warmer giant molecular clouds (Fig.\,\ref{gain}), where masing in principle could contribute to the line intensity at a significant rate. In fact, Tielens \& Hollenbach (unpublished) did notice this possibility in their models of the Orion PDR. We shall return to this point in a future paper. 

\subsection{Optically thick emission}

To explore the behaviour of the lines in the optically thick regime we need to resort to numerical calculations. To keep with the statistical character of this paper, we simplify things and exploit the Sobolev approximation using a Large Velocity Gradient (LVG) code. The basic parameter is the column density per radial velocity interval, $N({\rm O})/\delta v$. Once specified, the emergent intensity as the function of the gas kinetic temperature and density is calculated (diffuse background radiation due to, e.g., dust can also be included). 

\begin{figure}[t]
  \resizebox{\hsize}{!}{
  \rotatebox{00}{\includegraphics{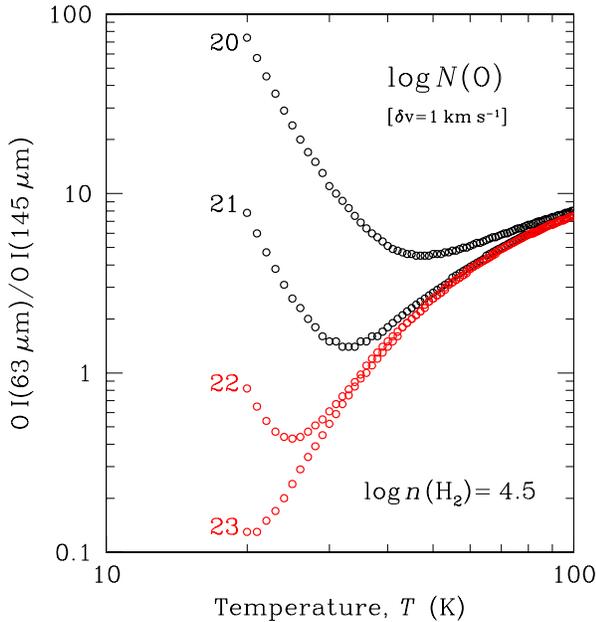}}
  }
  \caption{Numerical results for oxygen line ratios at characteristic interstellar cloud temperatures. Displayed are the column densities of atomic oxygen $N({\rm O})$, for $\delta v=1$\,\kms, needed to produce the line ratios for an \molh\ density of $3\times 10^4$\,\cmthree. For these parameters and at 20\,K, line center optical depths in this plot range from $\min \tau_0 = 0.05$ to $\max \tau_0 = 5 \times 10^5$ in the 145\,\um\ and in the 63\,\um\ lines, respectively. At 100\,K, these values are 26 and in excess of $4 \times 10^5$.
  }
  \label{column}
\end{figure}

Results for the average density of $3\times 10^4$\,\cmthree\ are shown in Fig.\,\ref{column}. On the 100\asec\ scale of the ISO observations, this density is already on the high side. As the temperature increases, all curves of Fig.\,\ref{column} tend toward the locus of the optically very thick ratio, which should be compared to Fig.\,\ref{theoratios} ($\tau \rightarrow \infty$). The recovery of this analytical result assures that the code behaves reliably also at these extremely high optical depths. 

From the figure it is evident, that in order to reduce the \oishort/\oilong\ line ratio to below 10, column densities exceeding $N({\rm O\,I})=10^{21}$\,\cmtwo\ would be required. Expressed as H column density or dust extinction, this would mean $N({\rm H})> 10^{24}$\,\cmtwo\ or \av\,$\gg 10^2$\,mag, rarely ever encountered on 100\asec\ scales.

Fig.\,\ref{column} also provides an explanation for the noticed paucity of data points for \oishort/\oilong\,\lapprox\,1 in Fig.\,\ref{obsratios}. Highly excessive column densities would be needed to populate this part of the diagram. However, taken together, the evidence speaks not in favour of the hypothesis of optically thick emission to explain the observed small oxygen line ratios. We consider next the case of 63\,\um\ self absorption due to a cold, relatively tenuous cloud component between the source and the observer.

\subsection{Foreground absorption}

At interstellar cloud temperatures, essentially all oxygen atoms reside in the lowest level of their ground state, with $J=2$. In terms of the center opacity $\tau_0$ of the $J=1 - 2$ \oishort\ line the column density of all atoms can then be approximated by 

\begin{equation}
N({\rm O\,I}) = 2 \times 10^{17} \,\tau_0\,\,\Delta v_{\rm fwhm}\hspace{0.5cm}{\rm cm}^{-2}
\end{equation}

where the line width is expressed in \kms. Parameter values of unity and an oxygen abundance $a({\rm O}) \le a({\rm O})_{\odot}=8.5 \times 10^{-4}$, then correspond to a hydrogen column of roughly $N({\rm H})\ge 2 \times 10^{20}$\,\cmtwo\ or, expressed differently, a visual extinction of \av\,\lapprox\,1\,mag for oxygen depletion factors less than five.

These numbers indicate that reasonably small amounts of cold foreground gas could potentially absorb in the 63\,\um\ line, leaving the 145\,\um\ line essentially unaffected and thereby altering the line ratios. Using the average values of the \oishort/\oilong\ ratios (Sect.\,3.2) would indicate that a reduction of the 63\,\um\ line by a factor of three ($2.9 \pm 1.6$) would be required to reconcile the observed low ratios with the average optically thin emission. This corresponds to average optical depths of order one to two ($\overline{\tau_0} = 1.0 ^{+0.5}_{-0.8}$). The profile of any {\it cold} foreground gas cloud is likely to be narrow, corresponding to $\Delta v_{\rm fwhm}\sim 1 {\rm ~to~} 2$\,\kms, say, whereas the line of the {\it warmer} background is likely broader (more turbulence, high velocity outflow gas), providing a quasi-continuum for the absorbing material \citep[see,~e.g.,][]{poglitsch1996}. Apart from being an ad hoc component, this seems to work fine. 

Conceptual difficulties with this scenario could arise for {\it nearby dark} clouds. These are cold, exhibiting barely any temperature contrast, and the lines are generally narrow. This should conceivably result in very little absorption, yet, for objects of this type small \oishort/\oilong\ ratios have been observed \citep{caux1999,liseau1999}. 

However, from detailed modeling of the observation with Odin\footnote{Odin is a Swedish-led satellite project funded jointly by the Swedish National Space Board (SNSB), the Canadian Space Agency (CSA), the National Technology Agency of Finland (Tekes) and Centre National d'Etude Spatiale (CNES).} of a nearby dark cloud in the ground state line of water, \water\ ($1_{10}-1_{01}$), \citet{larsson2005} found that an equivalent column density of $N({\rm H}_2)=6 \times 10^{20}$\,\cmtwo\ of 10\,K gas would be sufficient to account for the zero flux of the narrow self absorption in the \water\ line. For an oxygen abundance of $a({\rm O})=3.0 \times 10^{-4}$ \citep{savagesembach1996}, this translates to an $N({\rm O\,I})=1 \times 10^{17}$\,\cmtwo. Comparing with the numerical constant in Eq.\,3, this leads to an estimate of the optical depth of $\tau_0=2$ for the width of 1.1\,\kms. This would be sufficient to explain also the observed oxygen line ratios and we take this as independent support for the idea of foreground absorption of the 63\,\um\ line. 

However, the fact that the analysis of \oishort/\oilong\ data may require detailed modeling implies that their usefulness as diagnostics of the physical conditions in the emitting regions is significantly reduced.

\subsection{The \oishort/\ctwo\ intensity ratio}

We have shown that, for many sources, it is not the 145\,\um\ line which is relatively too strong, but most likely the 63\,\um\ line which is too weak. 

This would of course also affect the \oishort/\ctwo\ line ratio and could in part be the reason for the apparent distortion of the distribution to lower values (see Fig.\,\ref{obsratios}). 

An immediate implication would be that also the \oishort/\ctwo\ ratio would have its diagnostic power largely reduced.

\section{Conclusions}

Based on the examination of an extensive observational ISO-LWS material of [O\,I] and [C\,II] emission from outflows in star forming regions, we conclude the following:

\begin{itemize}
\item[$\bullet$] A large majority of observed intensity ratios \oishort/\oilong\ appears incompatible with optically thin emission, but could be reconciled with optically very thick line emission at relatively low temperatures.
\item[$\bullet$] Collisional excitation of the inverted levels of the ground state of O\,I by \molh\ can lead to inversion of the level populations. We show, however, that a masing 145\,\um\ line alone is likely not sufficient to explain the line ratio anomaly observed in dark clouds. 
\item[$\bullet$] Likewise, optically thick emssion in the lines would require unfeasibly large column densities to be consistent with the observations.
\item[$\bullet$] We provide evidence that foreground absorption, as has previously been advocated for the observed [O\,I] emission from distant luminous objects, is a viable explanation for observed oxygen line ratios also from dark clouds.
\item[$\bullet$] In galactic regions of outflows from young stellar objects, the fine structure lines of O\,I and C\,II have generally only limited diagnostic value.
\end{itemize}

\acknowledgements{This project has been supported by the Swedish National Space Board (SNSB). We thank Bengt Larsson and Fredrik Sch\"oier for valuable input to this paper.}

\appendix

\section{Statistical equilibrium}

In the steady state ($dn_i/dt=0$), and using common notation, the statistical equilibrium equations can be written in terms of the fractional population numbers $f_j = n_j/n_{\rm tot}$, as \citep[see, e.g., also][]{watson1980}

\begin{equation}
f_{j} \left [ \sum_{k} n_{\rm coll}\,q_{jk} + \sum_{k<j} A_{jk}\right ] -
\sum_{k} f_k\,n_{\rm coll}\,q_{kj} + \sum_{k>j} f_{k}\,A_{kj} = 0
\end{equation}
 
For the $^3P_J$ inverted ground state of O\,I with energies $E_2 < E_1 < E_0$, the exact analytical solution of this three level system is

\begin{equation}
f_0 = \frac{ \left ( a + b\,c \right ) d}{e\,c - g}
\end{equation}

\begin{equation}
f_1 = \frac{ \left ( a + b\,c \right ) e}{e\,c - b} - h
\end{equation}

\begin{equation}
f_2 = 1 - \left ( f_0 + f_1 \right )
\end{equation}

where

\begin{equation}
a = n_{\rm coll}\,q_{21}
\end{equation}

\begin{equation}
b = n_{\rm coll}\,q_{10} \left ( n_{\rm coll}\,q_{01} - A_{01} \right )
\end{equation}

\begin{equation}
c = n_{\rm coll}\left ( q_{10} + q_{12} \right ) + A_{12} 
\end{equation}

\begin{equation}
d = n_{\rm coll}\,q_{10}
\end{equation}

\begin{equation}
e = n_{\rm coll}\left ( q_{01} + q_{02}\right ) + A_{01} + A_{02} 
\end{equation}

\begin{equation}
g = n_{\rm coll}\,q_{10} \left ( n_{\rm coll}\,q_{01} + A_{01} \right )
\end{equation}

\begin{equation}
h = \frac{q_{20}}{q_{10}}
\end{equation}

Recently, the relevant atomic data have been compiled by \citet{schoier2005} in a convenient form and made readily accessible via the internet at \texttt{http://www.strw.leidenuniv.nl/$\sim$moldata/}

\end{document}